\documentclass[conference]{IEEEtran}

\usepackage{mathptmx}

\usepackage{graphicx}
\usepackage{amsmath}
\usepackage{amssymb}
\usepackage{booktabs}
\usepackage{tabularx}
\usepackage{cite}
\usepackage{url}
\usepackage[bookmarks=false,draft]{hyperref}
\usepackage{textcomp}
\usepackage{tikz}
\usetikzlibrary{positioning, arrows.meta, shapes.geometric, fit, backgrounds, calc, decorations.pathreplacing}

\begin{document}

\title{A Mission-Centric Cyber-Resilience Benchmark for Silent-Watch
       Operation of Electrified Ground-Platform Power Architectures}

\author{%
  \IEEEauthorblockN{Hongyu Wu and Raul Rodriguez}
  \IEEEauthorblockA{Department of Electrical and Computer Engineering\\Kansas State University, Manhattan, KS, USA, 66506\\
                    }
}

\maketitle

\begin{abstract}
Silent-watch operation makes electrified ground platforms depend on
supervisory energy management because mission loads must be sustained
from stored energy while the engine is off. This paper develops a
mission-centric cyber-resilience benchmark for this operating mode. The
benchmark connects battery state-of-charge (SOC) spoofing to mission outcomes rather
than evaluating the attack only through detector response or control
error. It combines a reduced-order DC-bus model, residual-based detection,
fallback shedding, and five mission-facing metrics for endurance, critical-load service,
priority-weighted loss-of-load cost, unsafe-voltage exposure, and
detection delay. The
study shows that SOC spoofing creates a structured stealth-versus-impact
envelope. Small biases have limited mission effect, intermediate biases produce an endurance deficit well approximated by
a first-order expression in bias magnitude, shed power, and average
battery draw, and large biases disable the
SOC-driven guard. The results also show that defense value depends on fallback depth,
not detection alone. An undersized fallback action can leave the
Defended case failing to complete the mission despite early detection.
MATLAB-to-Simulink parity across five regression scenarios provides a
software-verified basis for hardware-in-the-loop
testing.
\end{abstract}





\section{Introduction}

Electrification of military ground platforms changes the threat surface
that mission planners must reason about. A platform that previously
relied on mechanical drive and auxiliary generation may now depend on a
battery, a DC bus, and a supervisory controller that arbitrates between
mission-critical loads, sheddable loads, and protective actions. Silent
watch is a combat-vehicle energy-management problem in which the platform
must sustain mission loads from limited stored energy while the engine is
off~\cite{dattathreya2014silentwatch}. In that operating mode, the
supervisor becomes a consequential decision-maker because it determines
when to preserve energy, when to serve sheddable demand, and when to
protect the platform. If an adversary compromises the supervisor's
perception of state, the platform can lose endurance and critical-load
service even when the physical plant remains intact.

This paper combines mission-centric cyber resilience, vehicle
energy-management security, and cyber-physical testbed development. Mission-impact methods argue that cyber effects
should be evaluated by their consequences for mission objectives rather
than by intrusion counts or packet-level indicators alone~\cite{musman2010mission},
and mission-aware security work shows that asset criticality can depend
on mission phase and time~\cite{cam2013missionaware}. This framing is
appropriate for silent-watch operation because the operational question
is not only whether an intrusion is detected, but whether the platform
can still complete its assigned task. The Army Modernization Strategy
places ground combat vehicles within the Army's modernization
portfolio~\cite{usarmy2019modernization}, while prior combat-vehicle
work treats silent watch and energy management as operationally important
functions~\cite{dattathreya2014silentwatch}. What remains missing is a
compact benchmark that lets mission planners and defense developers
compare attacks and defenses on a common mission-facing basis.

Civilian vehicle and battery cybersecurity studies establish the
technical plausibility of the threat considered here, but they do not
provide the mission-mode benchmark needed for silent-watch analysis.
Prior surveys identify attack surfaces across vehicular communication,
sensing, charging, energy-management, and battery-control
layers~\cite{elrewini2020vehicular,aljohani2024evcyber,murlidharan2025bms,naseri2023cloudbms}.
Sensor-side and in-vehicle-bus studies show that spoofed or injected
measurements can affect vehicle cyber-physical behavior~\cite{elrewini2020sensors,kosmanos2020ids},
and connected or hybrid electric-vehicle studies show that adversarial
inputs can steer supervisory energy-management and powertrain-control
decisions~\cite{guo2021ems,guo2021hev}. These studies motivate the
attack model, but their focus is civilian vehicles, charging
systems, grid coupling, and powertrain control rather than the engine-off
supervisory loop of a tactical silent-watch platform.

Power-system cyber-physical security provides the closest technical
foundation for the measurement-integrity aspect of the problem. False
data injection formalizes how manipulated measurements can mislead
state-aware monitoring and control systems~\cite{liu2009fdia}. Moving
target defense studies show how changing system parameters or
communication assumptions can reduce an attacker's effectiveness
against power-grid cyber-physical systems~\cite{davis2012mtd,pappa2017mtd,liu2020dfacts}.
Learning-based detectors have been studied for false-data-injection
detection and localization~\cite{he2017dlfdia,wang2020locational}, but
our recent work shows that incomplete-network-information attacks can
defeat machine-learning detectors~\cite{liu2024tii}. This motivates the
present benchmark's use of a transparent residual detector in its first
version. A broader review of smart-grid cyber-physical attack and defense
appears in~\cite{zhang2021review}.

Real-time cyber-physical testbeds provide a path for translating the
benchmark beyond offline simulation. EV charging studies establish
grid-coupled vehicle cybersecurity as a relevant attack
surface~\cite{acharya2020pev,acharya2020charging}. OPAL-RT-class and
hardware-in-the-loop (HIL) co-simulation platforms have been used for
smart-grid performance and cyber-attack studies~\cite{bian2015opalrt,liu2018cosim},
EXata-coupled real-time simulation has been used to examine cyber-attack
impacts on microgrid controllers~\cite{pal2023exata}, and EV powertrain
HIL has been developed for cyber-physical security studies~\cite{yang2020evhil}.
Our prior Typhoon-HIL work further demonstrates real-time implementation
of CNN-based false-data-injection detectors~\cite{aminov2024kpec}. These
studies show that real-time deployment is feasible, but they do not
target the silent-watch supervisory loop or report cyber effects through
mission-facing quantities.

This paper addresses this gap by proposing a reduced-order,
architecture-level benchmark for silent-watch operation. The
contributions are as follows.
\begin{enumerate}
  \item We formulate a mission-centric cyber-resilience benchmark with
        five mission-facing metrics across three reference cases
        (Nominal, Attacked, and Defended) and two attack-window variants.
  \item We characterize a three-regime stealth-versus-impact envelope
        under SOC-bias sweeps and derive a first-order approximation of
        the endurance deficit for the delay regime.
  \item We identify a fallback-depth design constraint: undersized fallback shedding leaves the Defended case unable to complete the mission, while the mission-completion threshold lies between shed fractions of 0.2 and 0.4 in the reported sweep.
\end{enumerate}


The remainder of the paper is organized as follows.
Section~\ref{sec:models} defines the system, threat, and defense models.
Section~\ref{sec:methodology} describes the benchmark methodology and
mission metrics. Section~\ref{sec:results} presents the main results,
and Section~\ref{sec:discussion} provides discussion and conclusion.

\section{System and Threat Models}
\label{sec:models}

\subsection{Platform abstraction}

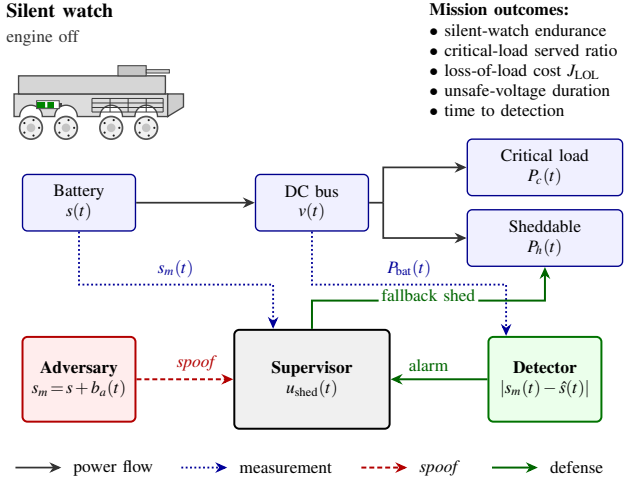
\begin{figure}[t]
    \centering
    \resizebox{0.95\linewidth}{!}{\begin{tikzpicture}[
    x=1cm, y=1cm,
    >={Stealth[length=1.7mm, width=1.5mm]},
    plant/.style={draw=blue!55!black, line width=0.55pt,
                  rounded corners=2pt, minimum height=8mm,
                  minimum width=16mm, align=center,
                  fill=blue!6, inner sep=2pt,
                  font=\scriptsize},
    super/.style={draw=black, line width=0.8pt,
                  rounded corners=2pt, minimum height=14mm,
                  minimum width=22mm, align=center,
                  fill=gray!12, inner sep=3pt,
                  font=\scriptsize\bfseries},
    threat/.style={draw=red!70!black, line width=0.8pt,
                   rounded corners=2pt, minimum height=12mm,
                   minimum width=16mm, align=center,
                   fill=red!7, inner sep=2.5pt,
                   font=\scriptsize},
    defense/.style={draw=green!45!black, line width=0.8pt,
                    rounded corners=2pt, minimum height=12mm,
                    minimum width=16mm, align=center,
                    fill=green!8, inner sep=2.5pt,
                    font=\scriptsize},
    flow/.style={->, line width=0.7pt, black!80},
    noflow/.style={line width=0.7pt, black!80},
    spoof/.style={->, line width=0.8pt, red!70!black, dashed,
                  dash pattern=on 2pt off 1.2pt},
    obs/.style={->, line width=0.75pt, blue!60!black,
                densely dotted},
    alarm/.style={->, line width=0.8pt, green!40!black},
    siglabel/.style={font=\scriptsize, inner sep=1.2pt,
                     fill=white, rounded corners=0.4pt}
  ]


  \begin{scope}[shift={(-4.22, 0.40)}, scale=0.80]
    \foreach \x in {0.30, 0.92, 1.72, 2.34} {
      \draw[fill=gray!11, draw=gray!65!black, line width=0.70pt]
        (\x, 0.20) circle (0.20);
      \fill[gray!46] (\x, 0.20) circle (0.076);
      \foreach \a in {0, 72, 144, 216, 288}
        \fill[gray!26!black]
          ($(\x, 0.20)+({0.144*cos(\a)},{0.144*sin(\a)})$) circle (0.019);
    }
    \draw[fill=gray!27, draw=gray!56!black, line width=0.55pt]
      (0.00, 0.40)
      -- (0.10, 0.40) arc(180:0:0.20)   
      -- (0.72, 0.40) arc(180:0:0.20)   
      -- (1.52, 0.40) arc(180:0:0.20)   
      -- (2.14, 0.40) arc(180:0:0.20)   
      -- (2.72, 0.40) -- (2.72, 0.80)
      -- (0.00, 0.80) -- cycle;
    \draw[fill=gray!38, draw=gray!56!black, line width=0.55pt]
      (2.72, 0.40) -- (3.00, 0.57) -- (3.00, 0.80) -- (2.72, 0.80) -- cycle;
    \draw[fill=gray!32, draw=gray!56!black, line width=0.55pt]
      (0.08, 0.80) rectangle (2.68, 1.12);
    \foreach \y in {0.48, 0.56, 0.64, 0.72}
      \draw[gray!44!black, line width=0.28pt] (1.38, \y) -- (2.64, \y);
    \foreach \xi in {1.38, 1.70, 2.02, 2.34, 2.64}
      \draw[gray!44!black, line width=0.28pt] (\xi, 0.48) -- (\xi, 0.72);
    \draw[fill=gray!44, draw=gray!60!black, line width=0.45pt]
      (1.84, 1.12) rectangle (2.34, 1.30);
    \draw[fill=gray!36, draw=gray!54!black, line width=0.38pt]
      (2.34, 1.17) rectangle (2.75, 1.23);
    \draw[fill=white, draw=gray!58!black, line width=0.40pt]
      (0.40, 0.54) rectangle (0.80, 0.68);
    \fill[gray!55!black] (0.80, 0.58) rectangle (0.85, 0.64);
    \fill[green!52!black] (0.42, 0.56) rectangle (0.71, 0.66);
    \draw[white, line width=0.9pt] (0.57, 0.56) -- (0.57, 0.66);
  \end{scope}

  \node[font=\footnotesize\bfseries, align=left, anchor=north west]
    at (-4.45, 2.45) {Silent watch};
  \node[font=\scriptsize, align=left, text=gray!40!black,
        anchor=north west]
    at (-4.45, 2.07) {engine off};

  \node[align=left, font=\scriptsize, anchor=north east] (out)
    at (4.45, 2.45)
    {\textbf{Mission outcomes:}\\[0.5pt]
     $\bullet$ silent-watch endurance\\
     $\bullet$ critical-load served ratio\\
     $\bullet$ loss-of-load cost $J_{\mathrm{LOL}}$\\
     $\bullet$ unsafe-voltage duration\\
     $\bullet$ time to detection};

  \node[plant] (bat)  at (-3.30,-0.50) {Battery\\$s(t)$};
  \node[plant] (bus)  at ( 0.00,-0.50) {DC bus\\$v(t)$};
  \node[plant, minimum width=22mm] (crit) at ( 3.30, 0.00)
    {Critical load\\$P_c(t)$};
  \node[plant, minimum width=22mm] (shed) at ( 3.30,-1.00)
    {Sheddable\\$P_h(t)$};

  \draw[flow] (bat.east) -- (bus.west);

  \coordinate (busbranch) at ($(bus.east)+(0.20,0)$);
  \draw[noflow] (bus.east) -- (busbranch);
  \draw[flow] (busbranch) |- (crit.west);
  \draw[flow] (busbranch) |- (shed.west);

  \node[threat]  (atk) at (-3.30, -3.00)
    {\textbf{Adversary}\\
     $s_m \!=\! s + b_a(t)$};
  \node[super]   (sup) at ( 0.00, -3.00)
    {Supervisor\\\mdseries$u_{\mathrm{shed}}(t)$};
  \node[defense] (det) at ( 3.30, -3.00)
    {\textbf{Detector}\\
     $|s_m(t) - \hat{s}(t)|$};

  \draw[obs] (bat.south) -- ++(0,-0.75)
    -| ($(sup.north)+(-0.55, 0)$);
  \node[siglabel, text=blue!60!black, anchor=south]
    at (-1.93, -1.60) {$s_m(t)$};

  \draw[obs] (bus.south) -- ++(0,-0.75)
    -| ($(det.north)+(-0.55, 0)$);
  \node[siglabel, text=blue!60!black, anchor=south]
    at (1.38, -1.60) {$P_{\mathrm{bat}}(t)$};

  \draw[spoof] (atk.east) -- (sup.west);
  \node[siglabel, text=red!70!black, font=\scriptsize\itshape,
        anchor=south] at (-1.65, -2.93) {spoof};

  \draw[alarm] (det.west) -- (sup.east);
  \node[siglabel, text=green!40!black, anchor=south]
    at (1.65, -2.93) {alarm};

  \draw[alarm] (sup.north) -- ++(0,0.40) -| (shed.south);
  \node[siglabel, text=green!40!black, anchor=south]
    at (1.65, -1.95) {fallback shed};

  \begin{scope}[shift={(0, -4.25)},
                every node/.style={font=\scriptsize}]
    \draw[flow]  (-4.20, 0) -- (-3.55, 0);
    \node[anchor=west] at (-3.50, 0) {power flow};
    \draw[obs]   (-1.85, 0) -- (-1.20, 0);
    \node[anchor=west] at (-1.15, 0) {measurement};
    \draw[spoof] ( 0.70, 0) -- ( 1.35, 0);
    \node[anchor=west, font=\scriptsize\itshape] at ( 1.40, 0) {spoof};
    \draw[alarm] ( 2.55, 0) -- ( 3.20, 0);
    \node[anchor=west] at ( 3.25, 0) {defense};
  \end{scope}

\end{tikzpicture}}
    \caption{Conceptual view of the silent-watch threat surface and the
           benchmark's evaluation loop. A ground platform with the
           engine off sustains critical loads from stored energy
           alone (top). The supervisor arbitrates between critical
           and sheddable demand on the basis of the measured state of
           charge $s_m(t)$. A cyber adversary biases $s_m$ during an
           attack window, which delays protective shedding and
           degrades mission endurance. A residual detector compares
           $s_m$ against an independently propagated estimate
           $\hat{s}(t)$ and triggers fallback shedding when the
           mismatch persists. The benchmark records five mission
           outcomes for every case.}
    \label{fig:overview}
\end{figure}

The platform is modeled as a DC-bus-centered source-storage-load-control
system, shown in Fig.~\ref{fig:overview}. The energy store is a
battery characterized by a capacity $E$ and an instantaneous state of
charge $s(t) \in [0, 1]$. An optional auxiliary source surrogate
$p_s(t)$ represents an APU or generator contribution and is held at
zero in the silent-watch case of record. The load is decomposed into a
critical demand $P_c(t)$ (sensors, communications, crew systems) and a
sheddable demand $P_h(t)$ (climate control, comfort loads). A
supervisory controller selects a shed fraction
$u_{\mathrm{shed}}(t) \in [0, 1]$ that determines what portion of
$P_h(t)$ is actually served. The battery balances any net demand left
after the source contribution. Bus voltage $v(t)$ is represented by a
low-order surrogate
\begin{equation}
  v_{\mathrm{eq}} = V_{\mathrm{nom}}
    - k_v \, \max(0, P_{\mathrm{net}})
    - k_s \, \max(0, 1 - s/s_{\mathrm{ref}}),
  \label{eq:vsurr}
\end{equation}
Bus voltage $v(t)$ tracks this equilibrium through the first-order lag
$\tau_v \dot{v}(t) = v_{\mathrm{eq}}(t) - v(t)$. The unsafe-voltage band is defined by $V_{\mathrm{min}}$, while mission
abort triggers if $s(t) \le s_{\mathrm{crit}}$ or $v(t) \le
V_{\mathrm{fail}}$, with $V_{\mathrm{fail}} < V_{\mathrm{min}}$.
Unless otherwise stated, all reported cases use the fixed parameter set listed in Table~\ref{tab:params}.

This abstraction is architecture-level. It is sufficient for
comparative benchmarking and remains easy to translate into Simulink
and HIL implementations without introducing modeling commitments that
the present scope does not justify. The numerical parameters are
illustrative and are not drawn from any specific platform's
controller; they are sized to a representative crewed ground vehicle
with a 24~kWh nominal usable battery and a 45-minute silent-watch
horizon for the headline study.

\begin{table}[t]
  \centering
  \caption{Benchmark parameters used in all reported runs.
           Identifiers match the symbols used in the text.}
  \label{tab:params}
  \footnotesize
  \begin{tabularx}{\columnwidth}{llX}
    \toprule
    Symbol & Value & Description \\
    \midrule
    $E$                       & 24 kWh    & nominal usable battery energy \\
    $s_0$                     & 0.80      & initial SOC \\
    $s_{\mathrm{warn}}$       & 0.65      & supervisor SOC warning \\
    $s_{\mathrm{crit}}$       & 0.20      & mission-fail SOC \\
    $V_{\mathrm{nom}}$        & 650 V     & nominal bus voltage \\
    $V_{\mathrm{warn}}$       & 590 V     & soft-shed bus threshold \\
    $V_{\mathrm{min}}$        & 585 V     & unsafe-band entry \\
    $V_{\mathrm{fail}}$       & 570 V     & mission-fail voltage \\
    $\tau_v$                  & 0.20 s    & bus first-order time constant \\
    $k_v$                     & 2 V/kW    & voltage droop slope \\
    $k_s$                     & 35 V      & low-SOC droop weight \\
    $s_{\mathrm{ref}}$        & 1.0       & SOC reference for droop activation \\
    $P_c$ baseline / step     & 10 / +2 kW & critical demand profile \\
    $P_h$ baseline / step     & 6 / +4 kW  & sheddable demand profile \\
    $\tau_d$                  & 0.08      & detector residual threshold \\
    confirmation window       & 5 s       & detector debounce time \\
    $\sigma_{s_m}$            & 0.015     & SOC sensor noise std.\ dev. \\
    $\sigma_{P_{\mathrm{bat}}}/P_{\mathrm{bat}}$ & 0.02 & current sensor noise (rel.) \\
    Mission horizon $T$       & 2700 s    & silent-watch evaluation window (45 min) \\
    Integrator step $\Delta t$ & 0.10 s    & fixed-step explicit Euler \\
    \bottomrule
  \end{tabularx}
\end{table}

\subsection{Supervisor}

The supervisor implements two protective paths that act on different
signals. The first is a soft-shed driven by the supervisor's view of
state of charge, $s_m(t)$, which becomes more aggressive as $s_m(t)$
approaches a warning threshold $s_{\mathrm{warn}}$ and a critical
threshold $s_{\mathrm{crit}}$. The second is a soft-shed driven by the
measured bus voltage, which engages when $v(t)$ falls toward its
unsafe band. In nominal operation $s_m(t) = s(t)$ and the two paths
reinforce one another. The distinction matters under attack because
the two paths have different exposure surfaces, as discussed below.

\subsection{Threat model}

The primary threat is state-of-charge spoofing at the supervisory
interface. The true SOC $s(t)$ is unchanged physically, but the
supervisor receives a biased measurement
\begin{equation}
  s_m(t) = \Pi_{[0,1]}\!\bigl(s(t) + b_a(t) + \eta_s(t)\bigr)
  \label{eq:spoof}
\end{equation}
during a specified attack window, where $\eta_s(t)$ is additive SOC
sensor noise ($\sigma_{s_m} = 0.015$) and $\Pi_{[0,1]}(\cdot)$ clips
the measurement to the physically admissible interval. The clipping
prevents the attack from delivering a physically impossible reading;
the bias delays protective shedding by keeping the controller-visible
SOC within the admissible range but above the true reserve. The
structure mirrors classical false-data-injection against state
estimation~\cite{liu2009fdia}. Battery-management threat surveys have
identified state-of-charge as a high-value manipulation target because
supervisory decisions depend on it directly~\cite{murlidharan2025bms}. Throughout this paper the bias is
positive, so the supervisor overestimates available energy, delays
protective action, and continues to support the sheddable load longer
than is appropriate for the actual reserve. We consider two
attack-window variants. The single-window variant applies the bias
over $[600, 2400]$~s. The multi-window variant applies the same bias
across two disjoint intervals, $[600, 1500]$~s and $[2100, 2700]$~s
(total 1500~s, versus 1800~s for the single-window case). This
variant tests whether the benchmark response changes when spoofing is
interrupted rather than continuous. Both variants share the headline
bias of $b_a = 0.50$ unless otherwise stated.

The benchmark quantifies the platform-level mission consequence without
attributing degradation to a specific delivery mechanism; coordinated
multi-channel spoofing is identified in Section~\ref{sec:discussion}
as future work.

\subsection{Defense}

The defense layer combines a residual detector with a conservative
fallback policy. A detector-side SOC predictor $\hat{s}(t)$ is
propagated from the measured battery-power balance and compared with
$s_m(t)$. When $|s_m(t) - \hat{s}(t)|$ exceeds a threshold $\tau_d$
for the confirmation interval, an alarm is raised and the fallback
controller reduces sheddable demand for the remainder of the mission.
We use a residual detector because it makes the benchmark
interpretable and separates mission impact from detector-family
choice. To avoid a noise-free detector setting, the predictor uses
noisy battery-current measurements (2\% relative Gaussian noise).

\section{Benchmark Methodology}
\label{sec:methodology}

\subsection{Reference cases}

Each scenario is exercised through three reference cases and two
attack-window variants:

\begin{itemize}
  \item \textbf{Nominal:} no attack, no defense. Establishes the
        energy-only mission envelope.
  \item \textbf{Attacked:} attack active, no defense. Establishes the
        unmitigated mission deficit.
  \item \textbf{Defended:} attack active, residual detector and
        fallback shed active. Establishes the mitigated mission
        envelope.
\end{itemize}

Note that, for sensitivity studies, the bias $b_a$ and the fallback shed fraction
are swept independently while all other parameters are held fixed.

\subsection{Mission metrics}

Five mission-facing metrics are proposed and recorded for every case:

\begin{itemize}
  \item \textbf{Silent-watch endurance} (minutes): the time to mission
        abort, defined as $s(t) \le s_{\mathrm{crit}}$ or
        $v(t) \le V_{\mathrm{fail}}$. Cases that reach the full horizon
        without abort are assigned the horizon duration.
  \item \textbf{Critical-load served ratio:} the fraction of integrated
        critical demand delivered over the mission horizon. After
        mission abort, remaining demand is treated as unserved.
  \item \textbf{Priority-weighted loss-of-load cost}
        ($J_{\mathrm{LOL}} \in [0,1]$): normalized unserved mission
        load weighted $w_c = 10$ (critical) and $w_h = 1$ (sheddable)
        over the fixed horizon $T$; after abort, all requested loads
        count as unserved for the remainder of $T$:
\begin{equation}
  J_{\mathrm{LOL}} = \frac{
    \int_0^{T}\bigl[w_c(P_c^{\mathrm{req}}-P_c^{\mathrm{srv}})^+
    + w_h(P_h^{\mathrm{req}}-P_h^{\mathrm{srv}})^+\bigr]\,dt}{
    \int_0^{T}\bigl[w_c P_c^{\mathrm{req}}
    + w_h P_h^{\mathrm{req}}\bigr]\,dt}
  \label{eq:jlol}
\end{equation}
where $(\cdot)^+ = \max(\cdot, 0)$.
  \item \textbf{Unsafe-voltage duration} (seconds): the time $v(t)$
        spends below $V_{\mathrm{min}}$ during the active mission
        interval.
  \item \textbf{Time to detection} (seconds): the elapsed time from
        attack onset to confirmed alarm. Reported as $\infty$ when no
        alarm is raised.
\end{itemize}

These five metrics are deliberately mission-facing rather than
control-theoretic. They allow platform stakeholders and control engineers to compare
cases on a common footing. The benchmark prioritizes critical-load
continuity; sheddable-load service is the resource sacrificed by
fallback control.



\subsection{Implementation and reproducibility}

The implementation is organized around traceable benchmark cases rather
than informal simulation runs. A MATLAB reference model defines the
offline benchmark, and a Simulink model preserves the same state-update
logic for later HIL execution. A regression harness runs both
implementations on five regression scenarios and compares the
mission-level outputs. This verifies the software translation before
OPAL-RT/EXataCPS deployment, so later discrepancies can be attributed to
HIL effects such as communication timing, I/O latency, or converter
dynamics.

\begin{figure}[t]
  \centering
  \includegraphics[width=\columnwidth]{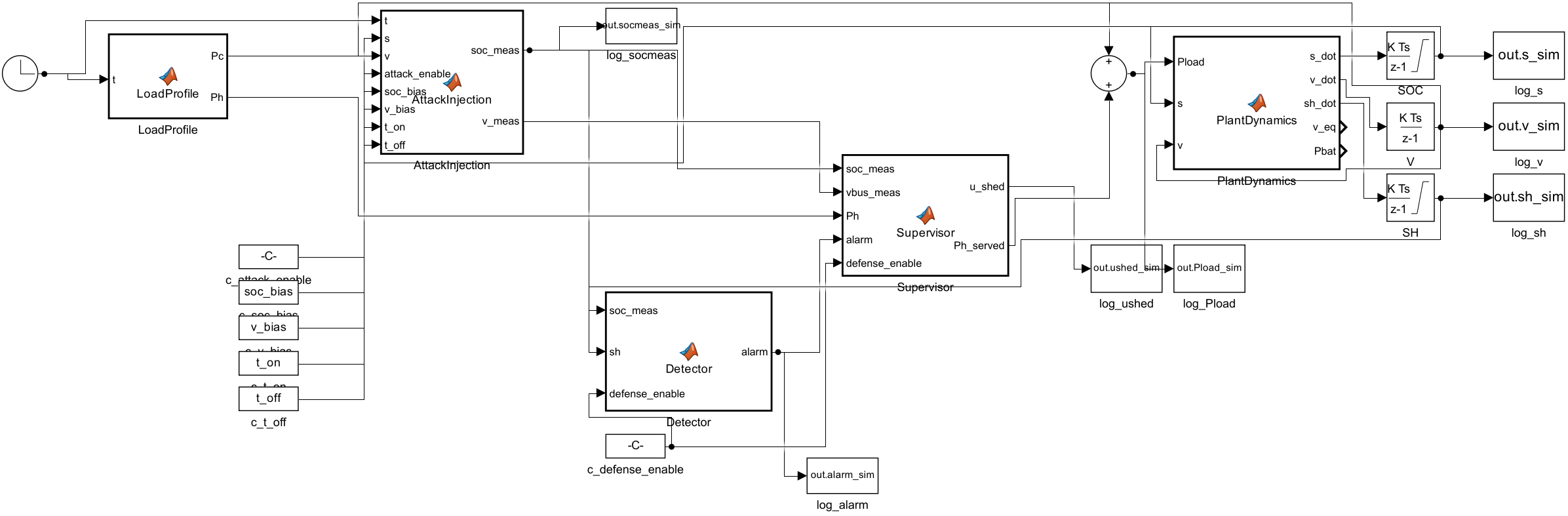}
  \caption{Simulink implementation of the benchmark. Plant, supervisor,
           attack, and detector-plus-fallback are separate
           subsystems that mirror the conceptual architecture of
           Fig.~\ref{fig:overview}.}
  \label{fig:simulink}
\end{figure}

\section{Results}
\label{sec:results}

\subsection{Headline three-case comparison}
\label{sec:results-headline}

Table~\ref{tab:headline} reports the five mission metrics for the
three reference cases under the headline parameter set ($b_a = 0.50$,
single-window attack on $[600, 2400]$~s, 45-minute silent-watch
horizon).

\begin{table}[t]
  \centering
  \caption{Headline three-case comparison, $b_a = 0.50$,
           single-window attack on $[600, 2400]$~s.}
  \label{tab:headline}
  \footnotesize
  \begin{tabular}{lrrrrr}
    \toprule
    Case      & End.\ (min) & Crit.\ ratio
              & $J_{\mathrm{LOL}}$ & Unsafe (s) & Det.\ (s) \\
    \midrule
    Nominal            & 45.00 & 1.000 & 0.021 & 0 & $\infty$ \\
    Attacked           & 42.90 & 0.952 & 0.053 & 0 & $\infty$ \\
    \textbf{Defended}  & \textbf{45.00} & \textbf{1.000} & \textbf{0.045} & \textbf{0} & \textbf{5} \\
    \bottomrule
  \end{tabular}
\end{table}

\begin{figure}[t]
  \centering
  \includegraphics[width=\columnwidth]{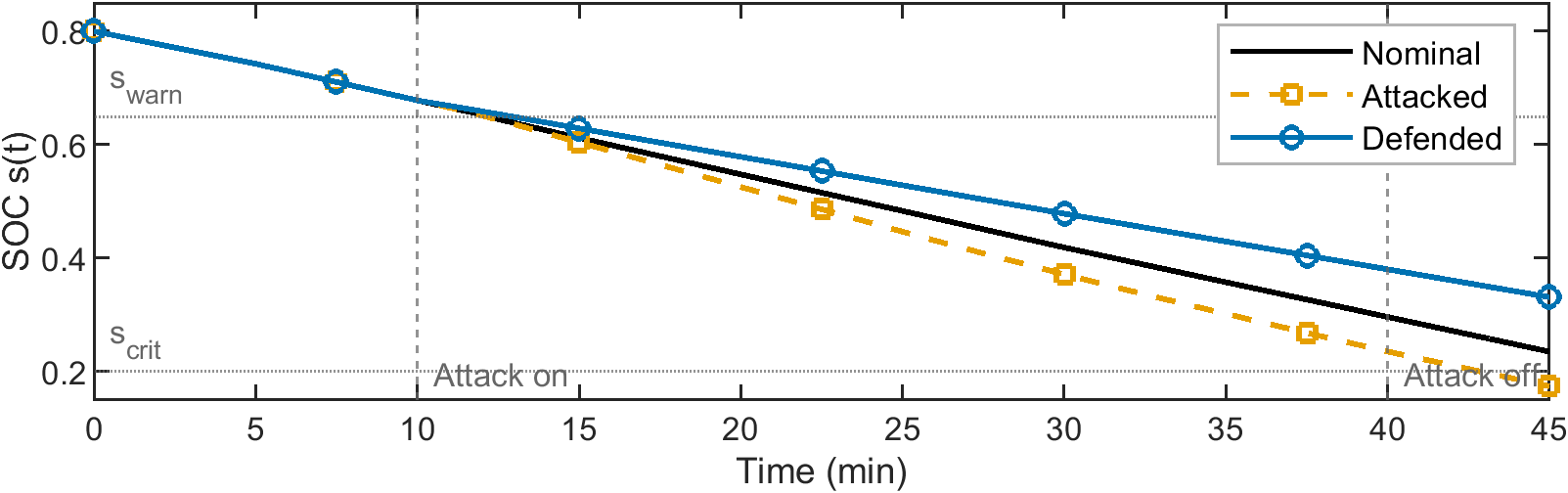}
  \caption{Battery state of charge $s(t)$ for the three reference
           cases. The Attacked case continues drawing on the
           sheddable load past the SOC warning threshold because the
           supervisor sees an inflated measurement.}
  \label{fig:soc}
\end{figure}

\begin{figure}[t]
  \centering
  \includegraphics[width=\columnwidth]{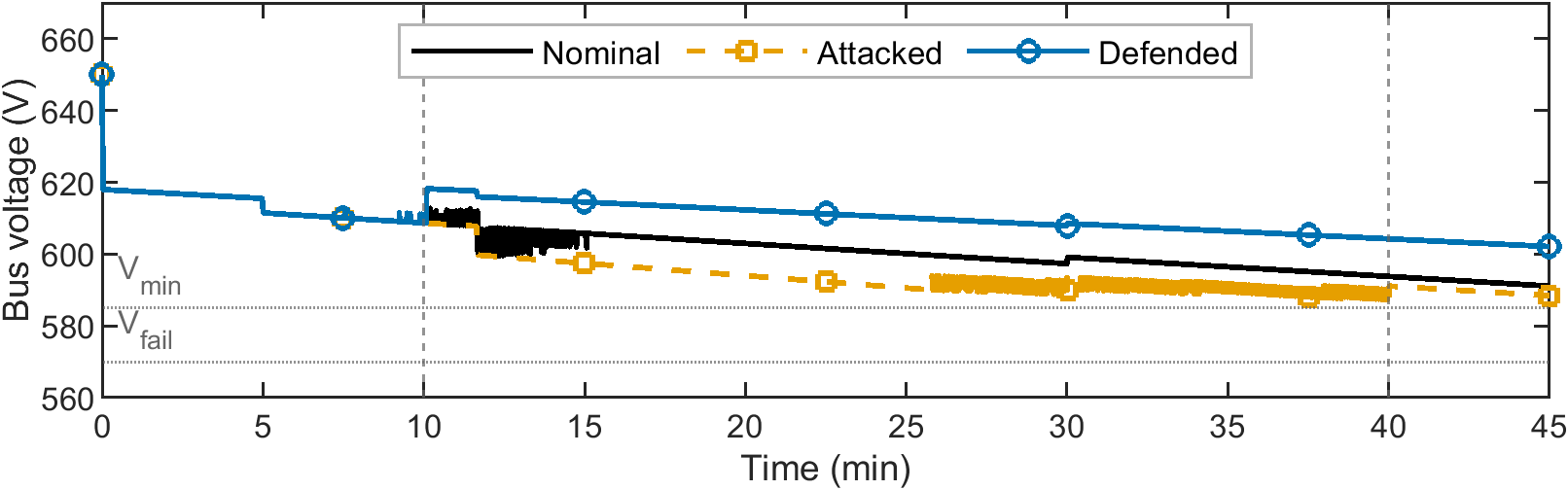}
  \caption{Bus voltage $v(t)$ for the three reference cases. The
           Attacked case shows a deeper voltage sag as delayed
           shedding allows higher battery draw, but the voltage-driven
           soft-shed path keeps all three cases above $V_{\mathrm{min}}$
           throughout the mission.}
  \label{fig:vbus}
\end{figure}

Under the 45-minute horizon, all three cases record
$T_{\mathrm{unsafe}} = 0$ because the voltage-based soft-shed path
engages before the bus crosses $V_{\mathrm{min}}$. The SOC-only
attack therefore manifests through energy-management decisions rather
than unsafe-voltage exposure; the metric is retained because it
becomes relevant under undersized fallback depth or future
voltage-channel attacks. In the Attacked case, the biased SOC
measurement delays protective shedding; the mission aborts at
42.90~min with a critical-load ratio of 0.952, yielding the highest
$J_{\mathrm{LOL}} = 0.053$. The Defended case detects the SOC
inconsistency after the 5-s confirmation window and enters fallback
shedding. It completes the 45-min horizon with full critical-load
service, but deliberate sheddable-load curtailment raises
$J_{\mathrm{LOL}} = 0.045$, above the Nominal value of 0.021 and
below the Attacked value. The defense trades lower-priority service
for critical-load continuity and mission completion.
Figs.~\ref{fig:soc} and~\ref{fig:vbus} show the underlying SOC and
bus-voltage trajectories.

\subsection{Stealth-versus-impact regime structure}
\label{sec:results-stealth}

\begin{figure}[t]
  \centering
  \includegraphics[width=\columnwidth]{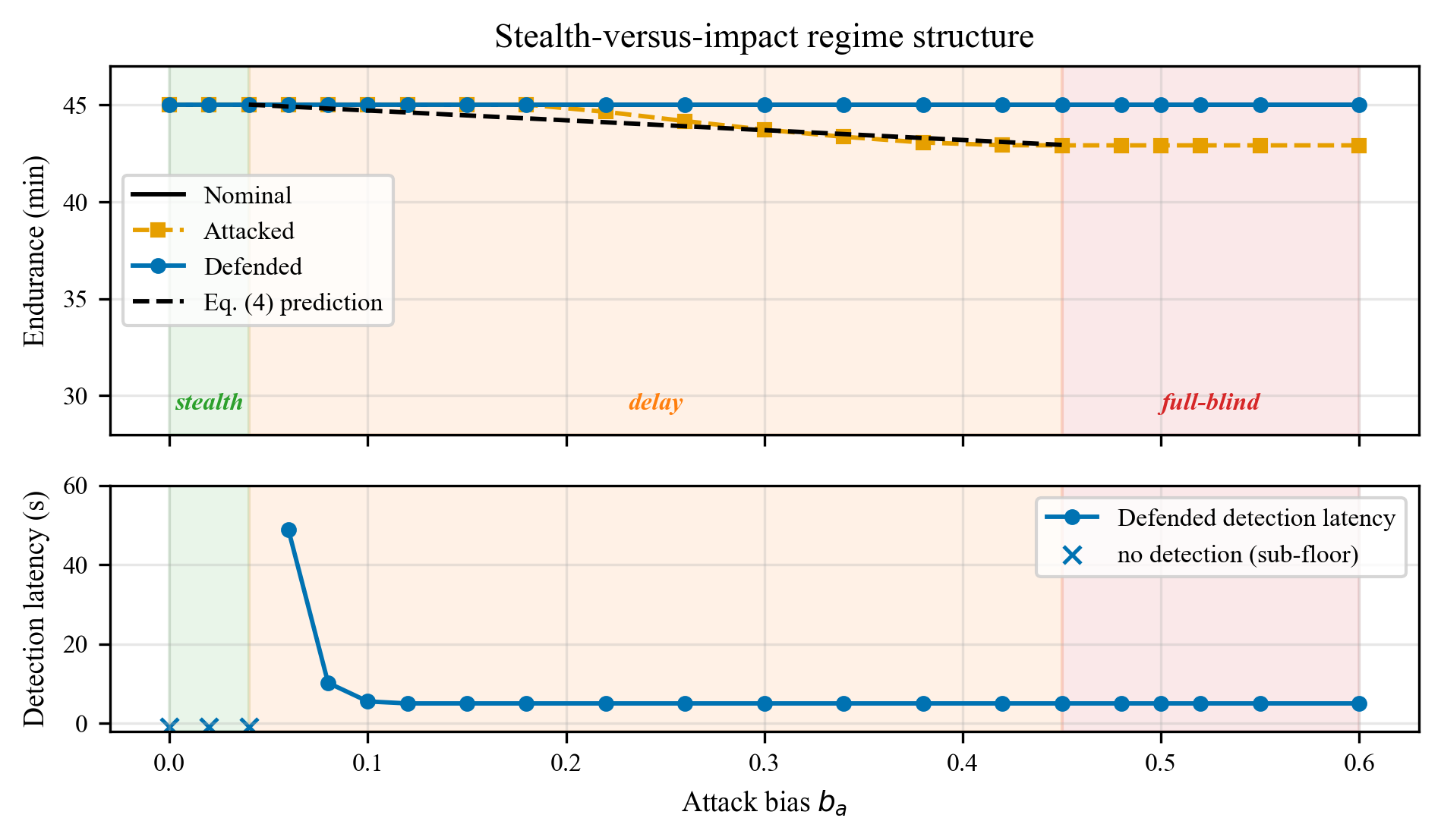}
  \caption{Stealth-versus-impact regime structure obtained by sweeping
           the SOC-spoofing bias $b_a$. Top panel: silent-watch
           endurance for the three reference cases. Bottom panel:
           detection latency for the Defended case. Background bands
           mark the stealth (left), delay (center), and full-blind
           (right) regimes.}
  \label{fig:bias_regimes}
\end{figure}

Fig.~\ref{fig:bias_regimes} reports the result of sweeping the bias
$b_a$ across $[0, 0.6]$ while holding all other parameters fixed, with the
three reference cases evaluated at each point. Three regimes are
visible. Below approximately $b_a = 0.04$ the residual remains under
the detector's noise floor, the alarm does not fire, and the induced
endurance loss in the Attacked case stays under 2\% of the Nominal
envelope. Between approximately $b_a = 0.04$ and $b_a = 0.45$ the
detector fires within tens of seconds of attack onset, and the
Defended case recovers to near-Nominal or higher endurance through
fallback shedding. Above $b_a \approx s_{\mathrm{warn}} - s_{\mathrm{crit}} = 0.45$ the
SOC-measurement guard is fully blinded; the undefended Attacked case
loses approximately 5\% of endurance relative to the Nominal case and
incurs a $J_{\mathrm{LOL}}$ more than twice that of the Nominal case;
the Defended case avoids both consequences through early fallback
shedding.

The regime boundaries follow from the SOC threshold structure. In the
delay regime, the biased SOC measurement postpones soft shedding until
the true SOC has fallen approximately $\Delta b = b_a - b_0$ below the
nominal shed point, where $b_0$ is the effective bias floor below which
the mission effect is negligible. Let $\bar{P}_{\mathrm{bat}}$ denote the average net battery draw and
$P_{\mathrm{shed}}$ denote the sheddable power $P_h$ that would have
been reduced at the soft-shed point under nominal operation. The shedding delay is then
$\Delta t_{\mathrm{delay}} \approx \Delta b \, E / \bar{P}_{\mathrm{bat}}$.
During this delay, the battery continues to serve $P_{\mathrm{shed}}$
beyond the post-shed load; the associated extra energy
$P_{\mathrm{shed}} \, \Delta t_{\mathrm{delay}}$ shortens the
remaining mission by approximately this extra energy divided by
$\bar{P}_{\mathrm{bat}}$. The endurance deficit therefore satisfies
\begin{equation}
  \Delta\mathrm{Endurance} \approx
    \frac{(b_a - b_0) \, P_{\mathrm{shed}} \, E}
         {\bar{P}_{\mathrm{bat}}^2},
  \label{eq:endbound}
\end{equation}
where $b_0 \approx 0.04$ is the effective bias floor and $E$ the
nominal usable battery capacity. When $E$ is in kWh and
$\bar{P}_{\mathrm{bat}}$, $P_{\mathrm{shed}}$ are in kW, the bound
carries units of hours. The predicted curve matches the
observed delay-regime deficit in Fig.~\ref{fig:bias_regimes} with
deviation below one minute. Under this threshold logic, the
full-blind regime begins when the bias exceeds
$s_{\mathrm{warn}} - s_{\mathrm{crit}}$ ($0.45$ here). The benchmark
therefore gives a compact analytical approximation of the attack
envelope rather than only pointwise simulation comparisons.



\subsection{Fallback-depth sensitivity}
\label{sec:results-depth}

A sweep of the fallback shed fraction from 0.2 to 1.0 under the
large-bias attack reveals that the defense is not uniformly
beneficial. Table~\ref{tab:depth} reports endurance and
$J_{\mathrm{LOL}}$ for the Defended case across the swept depth.

\begin{table}[t]
  \centering
  \caption{Fallback-depth sensitivity under the large-bias attack
           ($b_a = 0.50$, $T = 45$~min). Defended-case endurance and
           priority-weighted loss-of-load cost as functions of the
           post-alarm shed fraction.}
  \label{tab:depth}
  \begin{tabular}{ccr}
    \toprule
    Shed fraction & End.\ (min) & $J_{\mathrm{LOL}}$ \\
    \midrule
    0.2 & 44.54 & 0.022 \\
    0.4 & 45.00 & 0.023 \\
    0.6 & 45.00 & 0.034 \\
    0.8 & 45.00 & 0.045 \\
    1.0 & 45.00 & 0.056 \\
    \bottomrule
  \end{tabular}
\end{table}

At insufficient depth (0.2), the fallback action fails to sustain the
mission; the Defended case aborts at 44.54~min, only 1.6~min beyond
the undefended Attacked case (42.90~min). Detection alone, without
adequate shedding depth, does not recover the mission. Above a shed
fraction of 0.4, the defense completes the full 45-min horizon.
The $J_{\mathrm{LOL}}$ rises monotonically with shed fraction because
deeper shedding incurs more deliberate sheddable-load curtailment. At
the nominal operating depth of 0.8, $J_{\mathrm{LOL}} = 0.045$
remains below that of the undefended Attacked case (0.053). At depth
1.0, full curtailment of all sheddable load from alarm onward raises
$J_{\mathrm{LOL}}$ to 0.056, which narrowly exceeds the Attacked
baseline; this establishes an upper bound on beneficial shed depth. The
result confirms that early detection with adequate but not excessive
fallback depth is mission-beneficial. The result establishes a
minimum-viable-depth requirement as a transferable architectural
constraint rather than a point-tuned setting.

\begin{figure}[t]
  \centering
  \includegraphics[width=\columnwidth]{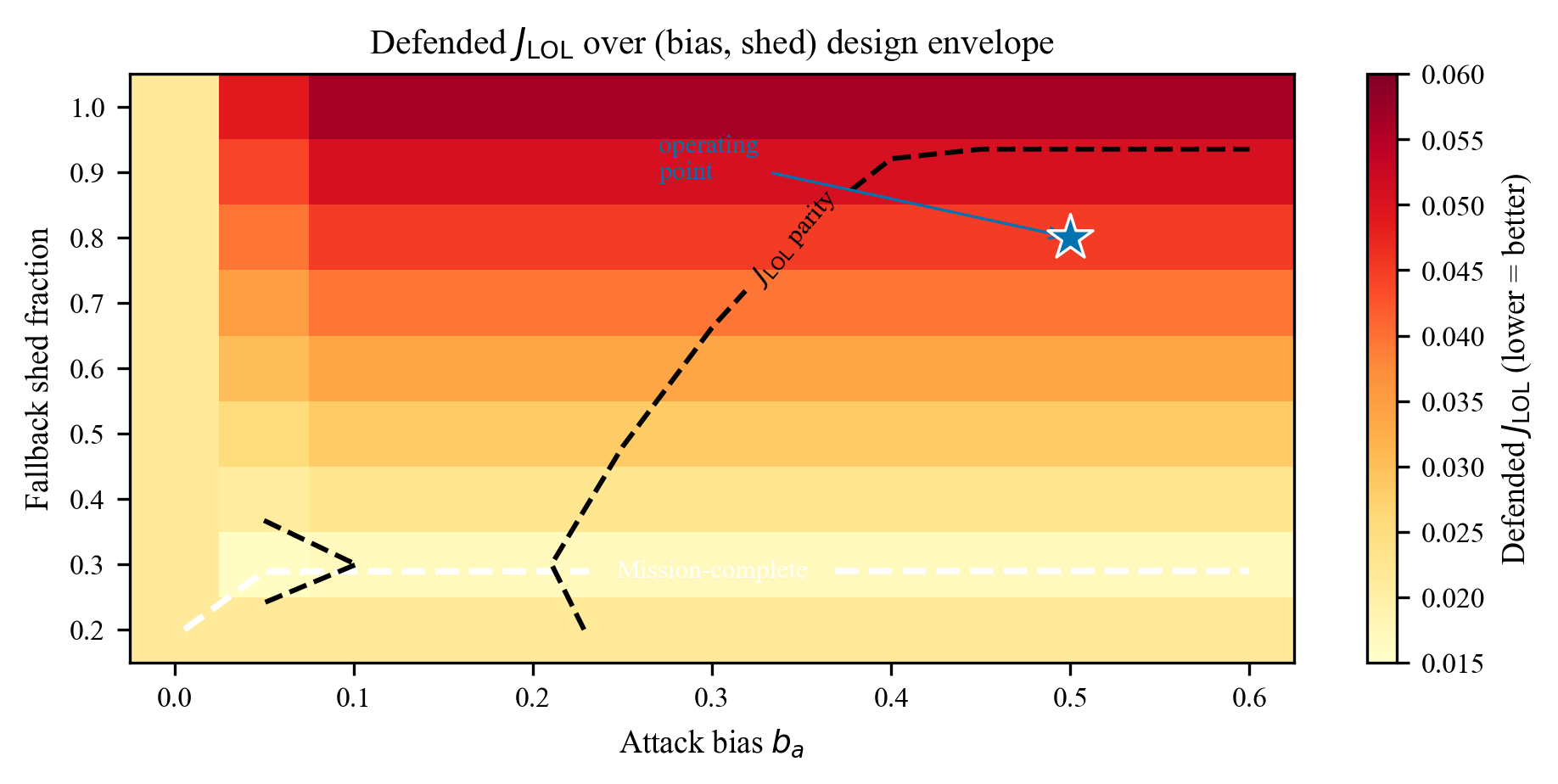}
  \caption{Defended $J_{\mathrm{LOL}}$ over the joint (bias,
           fallback-shed-fraction) design envelope. The white-dashed
           contour marks the mission-completion boundary (endurance
           $= 45$~min); the black-dashed contour marks the
           $J_{\mathrm{LOL}}$ parity boundary. The star marks the
           chosen operating point, which lies inside both bounds.}
  \label{fig:bias_shed_heatmap}
\end{figure}

The depth requirement generalizes naturally to the joint
$(b_a, u_{\mathrm{shed}}^{\mathrm{fb}})$ design space.
Fig.~\ref{fig:bias_shed_heatmap} maps Defended-case $J_{\mathrm{LOL}}$
over the (bias, shed-fraction) grid. Two contours bound the useful
operating region. The white-dashed contour marks the mission-completion
boundary (Defended endurance $= 45$~min); this boundary runs
approximately horizontally near a shed fraction of 0.25--0.30 and is
nearly independent of bias, confirming that a modest fixed shed depth
is sufficient across all tested attack strengths. The black-dashed
contour marks the $J_{\mathrm{LOL}}$ parity boundary, at which
Defended $J_{\mathrm{LOL}}$ equals the Attacked baseline; above this
curve, over-shedding erodes the cost advantage. The chosen operating
point $(b_a, u_{\mathrm{shed}}^{\mathrm{fb}}) = (0.50, 0.80)$ sits
inside both bounds. The value $u_{\mathrm{shed}}^{\mathrm{fb}} = 0.80$
is not the cost-minimizing depth; it is a conservative selection that
provides margin against model uncertainty and partial shed-actuator
failure for later HIL experiments.

\subsection{Preparation for HIL validation}

The multi-window attack produced the same qualitative mission ordering
as the single-window case and was used primarily as a regression
scenario for parity verification. The MATLAB reference and Simulink
implementation were exercised on five regression scenarios: Nominal,
Attacked, Defended, Attacked-multi, and Defended-multi. The two implementations use the same fixed step and
state-update order. Across these scenarios, the endurance, critical-load ratio, $J_{\mathrm{LOL}}$,
unsafe-voltage duration, and detection-latency metrics match exactly,
and the SOC and bus-voltage trajectory differences are zero to
reported precision. This verifies the software translation before later
OPAL-RT/EXataCPS experiments introduce communication timing, I/O
latency, and converter dynamics.

\section{Discussion and Conclusion}
\label{sec:discussion}

The benchmark is intended for comparative mission evaluation rather than
platform-specific endurance prediction. Because the reduced-order DC-bus
model is used under a fixed attack-defense setup, the reported quantities
should be interpreted as changes in endurance, critical-load service,
unsafe-voltage exposure, and detection delay, rather than as absolute claims about
a particular vehicle. Within that scope, the results show that SOC
spoofing can be mapped to mission-facing effects and that fallback depth
is an architectural design variable rather than a post-alarm implementation
detail.

The main findings are the structured attack envelope and the fallback-depth
requirement. The bias sweep separates stealth, delay, and full-blind
regimes. Small biases have limited mission effect, intermediate biases
follow a closed-form endurance bound, and large biases disable the
SOC-driven guard. The fallback-depth sweep shows that detection alone is insufficient:
undersized shedding leaves the Defended case unable to complete the
mission despite early detection, and excessive shedding can raise
$J_{\mathrm{LOL}}$ above the Attacked baseline.

The MATLAB-to-Simulink parity result provides the software-verified
basis for OPAL-RT/EXataCPS HIL evaluation. Future work will use this
benchmark as a fixed baseline to test whether the attack-regime
structure and fallback-depth requirement persist under higher-fidelity
dynamics, communication timing effects, learning-based detection, and
coordinated multi-channel adversaries.

\bibliographystyle{IEEEtran}
\bibliography{references}

\end{document}